**FURTHER INVESTIGATIONS OF DUST GRAIN ORBITAL BEHAVIOR AROUND NEPTUNE.** R. Nazzario and T. W. Hyde, Center for Astrophysics, Space Physics and Engineering Research, Baylor University, Waco, TX 76798-7316, USA, phone: 254-710-2511 (email: Truell_Hyde@Baylor.edu).

**Introduction:** The time-scale for the formation lifetime of Neptune's rings is on the order of millions of years. Most current theories suggest the dust in these rings is a result of collisional processes [1-3]. This paper advances current investigations by tracking the orbital behavior of individual dust grains around Neptune while under the influence of a detailed magnetic field model. Dust grain orbital evolution is compiled for dust ejected from one of the three inner moons with comparisons drawn to previous efforts [1-10].

**Model:** It is reasonable to assume that impactors on Neptune's moons eject dust particles in much the same manner as previously observed for the moons of Jupiter and Earth. For Jupiter, such ejected dust particles have both a size and velocity distribution, which has not yet been measured and as such, is not well known. As a result, this work tracks the orbits for individual ejecta, assuming an initial random velocity for each particle. The size of each dust grain is also randomly selected to fall in the regime between 7 and 7000 μm in radius. The orbital parameters for each grain are calculated while taking into account Neptune's gravitational force (where both the spherical and oblate field expansion terms are considered), the solar radiation pressure force, the solar gravitational force, the gravitational fields produced by Neptune's moons, and (for charged grains) the interaction with the magnetic field of Neptune. This allows a determination of the dust particle's dynamics using a force equation for the particle, given as

$$\vec{F} = -\frac{GM_n m_d}{r^2}\hat{r} - \frac{3GM_n m_d J_2 R_n^2}{r^4}\left[\left(3\sin^2\theta - 2\right)\hat{r} - \sin 2\theta \hat{\theta}\right] + \frac{\beta GM_{sun} m_d}{R^2}\hat{R} + M_{sun}G\left[-\frac{\vec{r}}{R^3} + \vec{\rho}\left(\frac{1}{R^3} - \frac{1}{\rho^3}\right)\right] + \frac{3\varepsilon_0 \phi}{\rho s^2}\vec{v}\times\vec{B} \quad (1)$$

In Eq. (1), G is the gravitational constant, $M_n$ is the mass of Neptune, $m_d$ is the mass of the dust grain, r is the distance from the center of Neptune to the position of the particle, $\hat{r}$ is the unit vector from Neptune to the particle, $M_{sun}$ is the mass of the Sun, R is the distance from the Sun to the particle, and $\hat{R}$ is the unit vector from the Sun to the dust particle for the radiation pressure force. In the second to last term, $\vec{\rho}$ is the position vector from the Sun to Neptune [11].

The first term on the right hand side of Eq. (1) represents the gravitational attraction on the grain assuming a spherical Neptune [12]. The second term corrects this to include an oblateness term ($J_2$) for the Neptunian gravitational field where θ is the angle measured from Neptune's rotational axis. The third term describes the radiation pressure force due to the solar radiation incident upon the particle with β defined as the ratio of the solar radiation pressure force to the solar gravitational force [13]. β is calculated using

$$\beta = \frac{.6Q}{a\rho} \quad (2)$$

where Q is the radiation pressure efficiency (taken to be 1.0 for particles with radii larger than 1 μm), a is the radius of the particle (between 1 μm and 100 μm as specified earlier), and ρ is the density of the particle (assumed to be 1.4 g/cm$^3$). The final term in equation (1) takes into account the interaction of the magnetic field on a charged particle where $\phi$ is the potential of the grain, $\vec{B}$ is the magnetic field of Neptune, $\vec{v}$ is the velocity of the dust grain relative to the magnetic field and $\varepsilon_0$ is the permittivity of free space. The magnetic field ($\vec{B}$) is calculated using a model developed by Holme and Bloxham [14] where

$$\vec{B} = -\nabla\Phi \quad (3)$$

and the potential is defined by

$$\Phi = a\sum_{l=1}^{\infty}\left(\frac{a}{r}\right)^{l+1}\sum_{m=0}^{l}P_l^m(\cos\theta)\begin{pmatrix}g_l^m\cos(m\phi)\\+h_l^m\sin(m\phi)\end{pmatrix} \quad (4)$$

To insure accurate results, a planetary shadowing effect was taken into account whenever the particle moved behind Neptune. Additionally, the planet's orbit was taken to be elliptical in order to achieve a more realistic radiation pressure calculation. A fifth order Runge-Kutta method was employed for all numerical simulations.

**Results:** Each simulation initially assumed a random velocity of up to 500 m/s for the dust particle as measured relative to the parent body. This velocity distribution was selected due to experimental hypervelocity impact data showing that most such ejecta is created by low impact velocities. The initial distance from the surface of the grain's progenitor (Naiad, Thalassa or Despina) is taken to be 1000 m. The grains are launched into the shadow of Neptune with the Sun, Neptune and its moons initially aligned along the same axis. The potential on each particle was varied between –3.0 V (to simulate grains exposed to solar radiation) and -18.0 V (for particle's in Neptune's shadow). A size range of particles was studied using $\beta$ values from 0.001 to 0.1 with a step size of 0.001. Particle ejection speeds of 5.0, 50.0 and 500.0 m/s were assumed. Nine random impacts were simulated on each moon (for a total of 2700 particles) with each simulation allowed to run for 10 years.

**Conclusions:** In all cases, the orbital motion of the dust particles followed extremely complicated paths. As can be seen in Figure (1), for grains released from Naiad particles across the size regime (all ejection speeds) survived for the entire ten-year period. In the case of Thalassa (Figure 2), 1309 of 2700 grains released survived the ten-year period with all ejection ranges represented. For particles released from Despina (Figure 3), only those grains with initial velocities of 500 m/s or greater survived for more than 1 year. All others quickly impacted Despina. Discussion of the above results (along with several additional data sets) will be presented. Correlations with the orbital properties of the known rings of Neptune will also be explored.

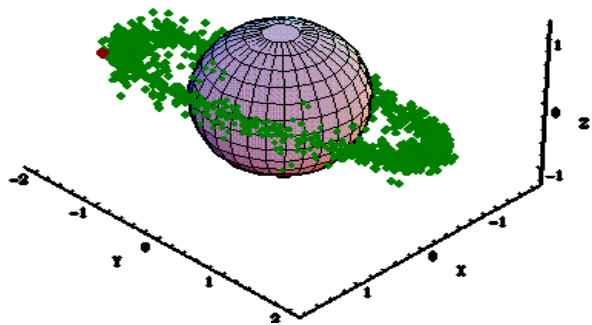

Figure 1. Particle distribution after 10 years for grains launched from Naiad.

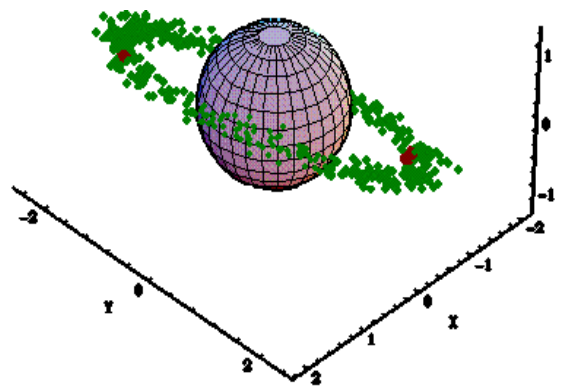

Figure 3. Particle distribution after 10 years for grains launched from Despina.

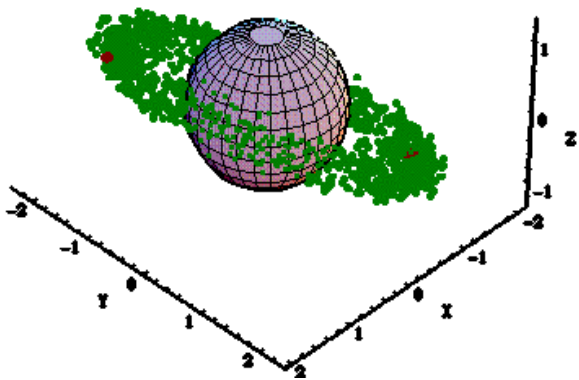

Figure 2. Particle distribution after 10 years for grains launched from Thalassa.